\documentclass[fleqn,usenatbib]{mnras}

\usepackage{newtxtext,newtxmath}
\usepackage[T1]{fontenc}

\DeclareRobustCommand{\VAN}[3]{#2}
\let\VANthebibliography\thebibliography
\def\thebibliography{\DeclareRobustCommand{\VAN}[3]{##3}\VANthebibliography}

\usepackage{graphicx}	% Including figure files
\usepackage{amsmath}	% Advanced maths commands
\usepackage{subcaption} % Shin added
\title[Four FRBs Discovered]{Four New Fast Radio Bursts Discovered in the Parkes 70-cm Pulsar Survey Archive}

\author[F. Crawford et al.]{
F. Crawford,$^{1}$\thanks{E-mail: fcrawfor@fandm.edu}\href{https://orcid.org/0000-0002-2578-0360}{\includegraphics[scale=0.08]{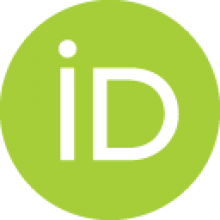}}
S. Hisano,$^{2}$\href{https://orcid.org/0000-0002-7700-3379}{\includegraphics[scale=0.08]{Orcid-ID.png}}
M. Golden,$^{1}$
T. Kikunaga,$^{2}$\href{https://orcid.org/0000-0002-5016-3567}{\includegraphics[scale=0.08]{Orcid-ID.png}}
A. Laity,$^{1,3}$
and D. Zoeller,$^{1}$ 
\\
$^{1}$Department of Physics and Astronomy, Franklin and Marshall College, P.O. Box 3003, Lancaster, PA 17604, USA\\
$^{2}$Kumamoto University, Graduate School of Science and Technology, Kumamoto, 860-8555, Japan\\
$^{3}$Department of Physics, Millersville University, P.O. Box 1002, Millersville, PA 17551, USA
}

\date{Accepted XXX. Received YYY; in original form ZZZ}

\pubyear{2022}

\begin{document}
\label{firstpage}
\pagerange{\pageref{firstpage}--\pageref{lastpage}}
\maketitle

\begin{abstract}
We present four new fast radio bursts discovered in a search of the Parkes 70-cm pulsar survey data archive for dispersed single pulses and bursts. We searched dispersion measures (DMs) ranging between 0 to 5000 pc cm$^{-3}$ with the HEIMDALL and FETCH detection and classification algorithms. All four of the FRBs discovered have significantly larger widths ($> 50$ ms) than almost all of the FRBs detected and cataloged to date. The large pulse widths are not dominated by interstellar scattering or dispersive smearing within channels. One of the FRBs has a DM of 3338 pc cm$^{3}$, the largest measured for any FRB to date. These are also the first FRBs detected by any radio telescope so far, predating the Lorimer Burst by almost a decade. Our results suggest that pulsar survey archives remain important sources of previously undetected FRBs and that searches for FRBs on time scales extending beyond $\sim 100$ ms may reveal the presence of a larger population of wide-pulse FRBs.
\end{abstract}



\section{Introduction and Background}

The Parkes 70-cm pulsar survey was conducted in the early 1990s with the Parkes 64-m radio telescope, and it covered the full southern sky visible from Parkes. A total of 44299 separate sky locations (beams) were gridded for the total planned survey. 
Survey observations were conducted between 1991 and 1994, and a total of 43842 beams (representing 99\% of the planned survey coverage) were observed.
Observations were conducted at a center frequency of 436 MHz, with a total bandwidth of 32 MHz split into 256 frequency channels (0.125 MHz per channel). Each channel was 1-bit sampled every 0.3 ms, and nominal integration times were 157 s per observation.  A search for periodicity candidates was performed in the original survey analysis, with a DM range extending from 0 to 777 pc cm$^{-3}$ or to the maximum expected Galactic DM from the NE2001 Galactic electron model \citep{cl02}, whichever was smaller. A total of 298 pulsars were detected in the survey, of which 101 were new discoveries (including 17 millisecond pulsars). The complete details and results from the original survey were presented in three papers \citep{mld+96,lml+98,dsb+98}.

The discovery of rotating radio transients (RRATs) in 2006 \citep{mll+06} and the first extragalactic fast radio burst (FRB) in 2007 (the Lorimer Burst; \citealt{lbm+07}) indicated that searches for dispersed impulsive signatures in pulsar surveys could reveal previously undetected astrophysical signals. Since then, the field of FRB science has rapidly grown, and a summary of the current state of the field can be found in recent reviews by \citet{phl+19,phl+22}. We have undertaken a re-analysis of the Parkes 70-cm pulsar survey in order to look for previously undetected dispersed single pulse events. A similar analysis has already been conducted on the large amount of data acquired with the multibeam receiver at Parkes between 1997 and 2001 \citep{zhr+20,yzw+21}.

\section{Survey Data Analysis}

For our analysis, we downloaded the full Parkes 70-cm data archive from the CSIRO data portal\footnote{\url{https://data.csiro.au/}}. We first converted the PSRFITS format files into filterbank format files for processing. The data were also converted from 1-bit to 8-bit samples to interface with the single pulse search software packages. Each separate observation was processed by HEIMDALL \citep{b12}\footnote{\url{https://sourceforge.net/projects/heimdall-astro}}, with a DM search range of 0 to 5000 pc cm$^{-3}$ to look for single pulse events. Boxcar matched filtering windows having integer powers of two samples ranging from 1 to 512 samples were applied to each dedispersed time series to maintain maximum sensitivity to pulse widths up to $\sim 150$ ms.

The resulting pulse detections from HEIMDALL were then analyzed by FETCH \citep{aab+20}\footnote{\url{https://github.com/devanshkv/fetch}} to determine the likelihood of a detected signal being real using pulse morphology metrics. FETCH rated each event with a likelihood probability of being real between 0 and 1. Every candidate with probability greater than 0.5 was inspected visually. Events that appeared realistic were then checked against known pulsars from past and current surveys \citep{k22}\footnote{\url{https://pulsar.cgca-hub.org}} for the presence of any pulsars in the vicinity that could have produced single pulses near the same DM.

\section{Results and Discussion}

A total of 719905 single pulse candidates were detected by HEIMDALL in the survey, of which 75774 were classified by FETCH as possibly real (probability $p \ge 0.5$). Note that 25\% of the FETCH classifications above 0.5 had $p \ge 0.9999$, indicating almost certainly real signals. 

All but 7 of the classified signals were either rejected by eye as not morphologically realistic or were determined to have come from known pulsars, some of which emitted many detectable pulses in a single observation (e.g., Vela and PSR J0437$-$4715 both appeared often in multiple survey beams in the vicinity of these pulsars). The fact that all but a handful of these signals were associated with known sources after checking the pulsar catalog illustrates the large number of known single pulse emitters present in the survey data. Of the 7 unidentified signals, three were weak impulses (S/N $\la 10$) with narrow widths ($< 4$ ms) and small (Galactic) DMs, indicating likely RRATs. However, none of these 3 warranted a clear and obvious claim of detection, and so they remain as possible detections that we do not report upon or discuss here.

\begin{figure}
\centering
\begin{subfigure}[t]{0.45\columnwidth}
    \centering
    \vspace{-2mm}
    \caption{FRB910730}
    \includegraphics[width=\textwidth]{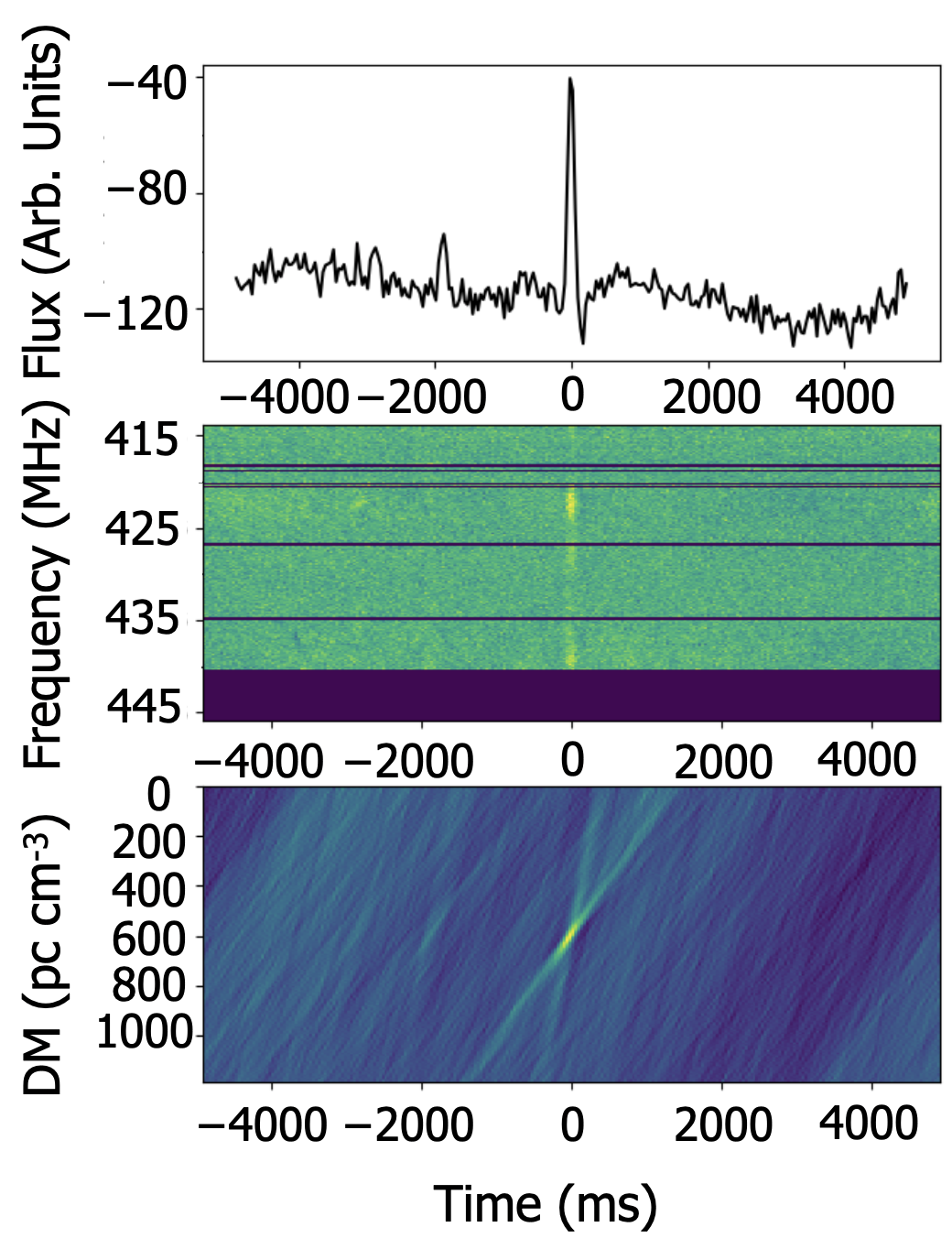}  
\end{subfigure}
~ %add desired spacing between images, e. g. ~, \quad, \qquad, \hfill etc. 
\begin{subfigure}[t]{0.46\columnwidth}
    \centering
    \vspace{-2mm}
    \caption{FRB920428}
    \includegraphics[width=\textwidth]{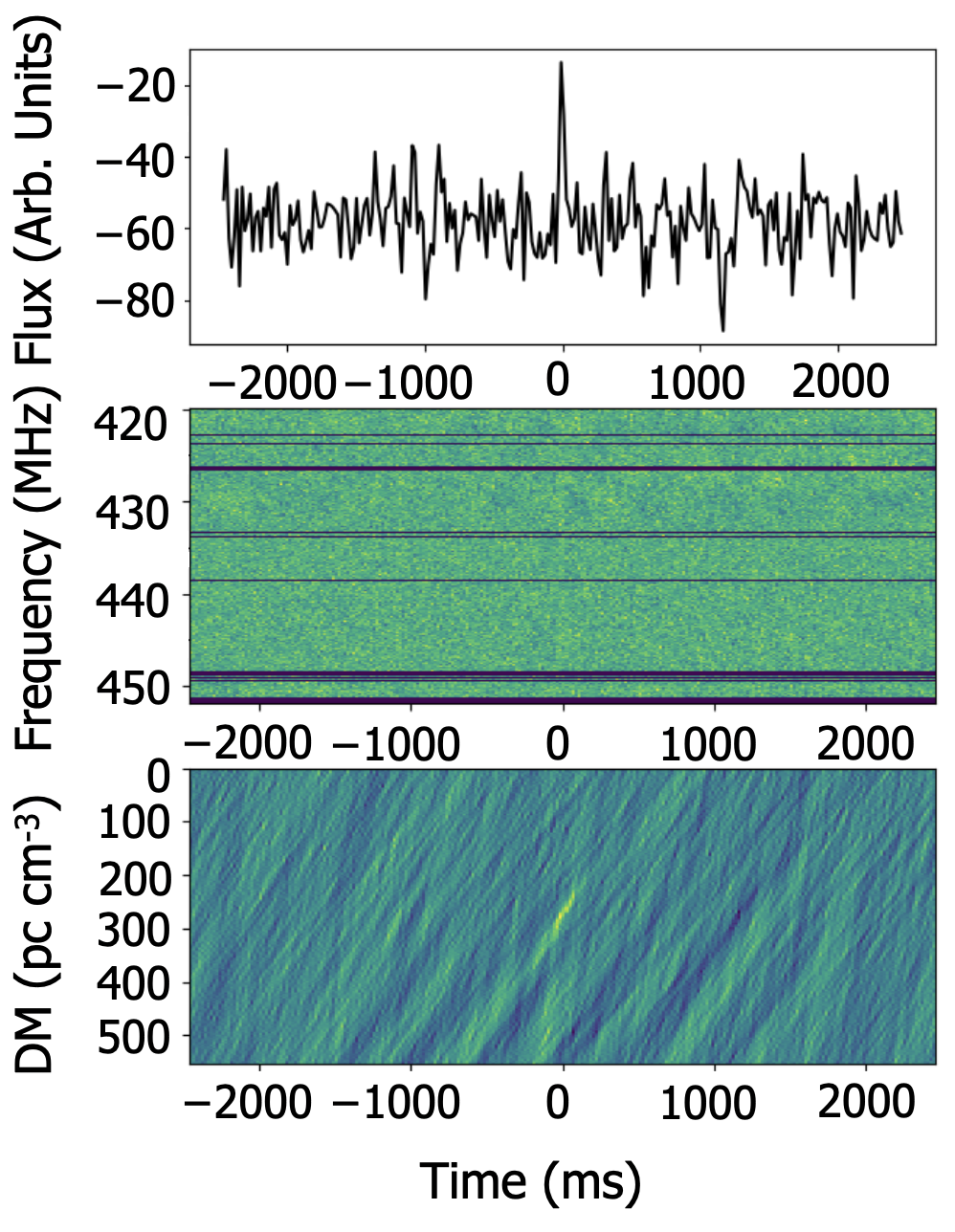}
\end{subfigure}
~ %add desired spacing between images, e. g. ~, \quad, \qquad, \hfill etc. 
\begin{subfigure}[t]{0.47\columnwidth}
    \centering
    \vspace{4.6mm}
    \caption{FRB920913}
    \includegraphics[width=\textwidth]{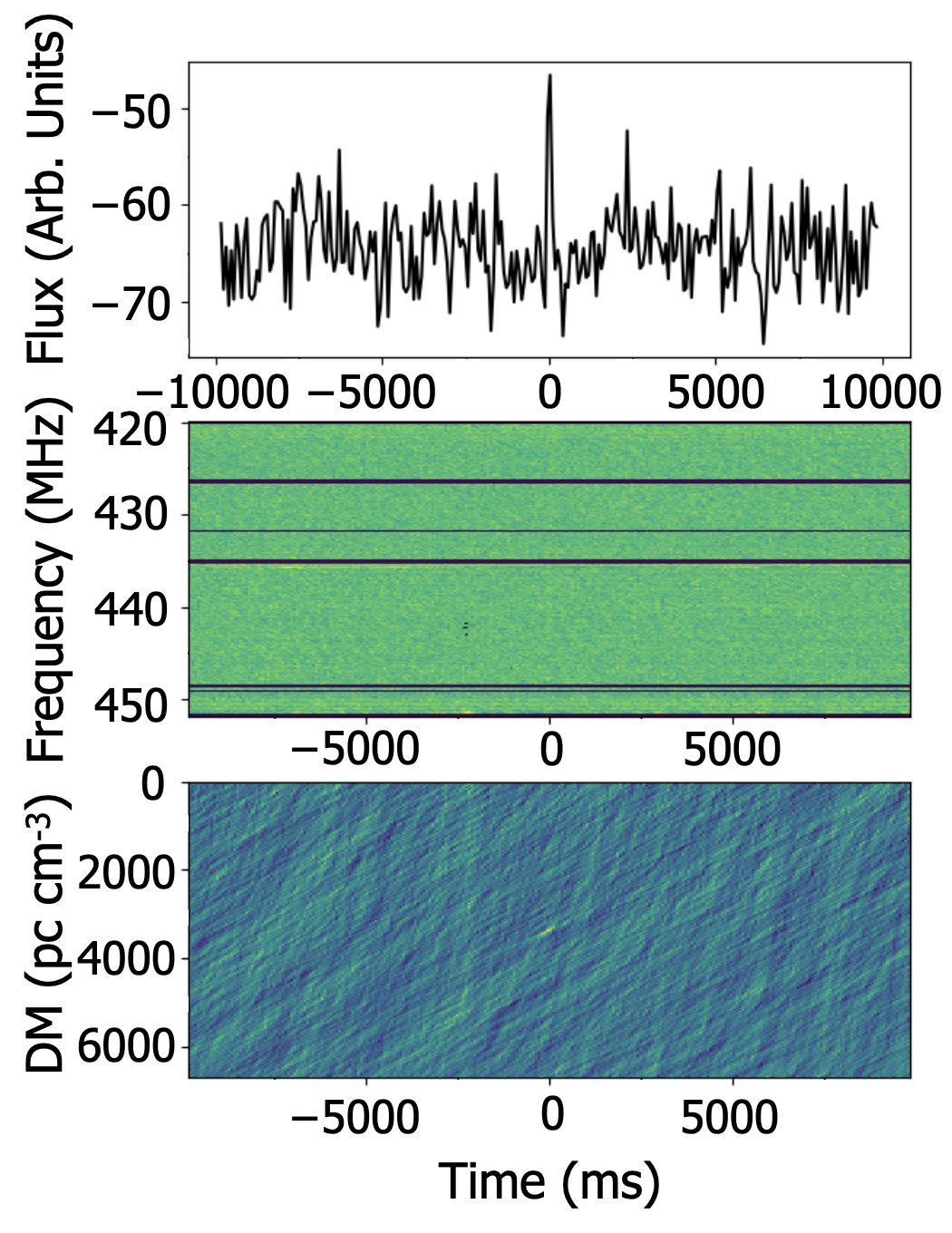}
\end{subfigure}
~ %add desired spacing between images, e. g. ~, \quad, \qquad, \hfill etc. 
\begin{subfigure}[t]{0.49\columnwidth}
    \centering
    \vspace{4mm}
    \caption{FRB921212}
    \includegraphics[width=\textwidth]{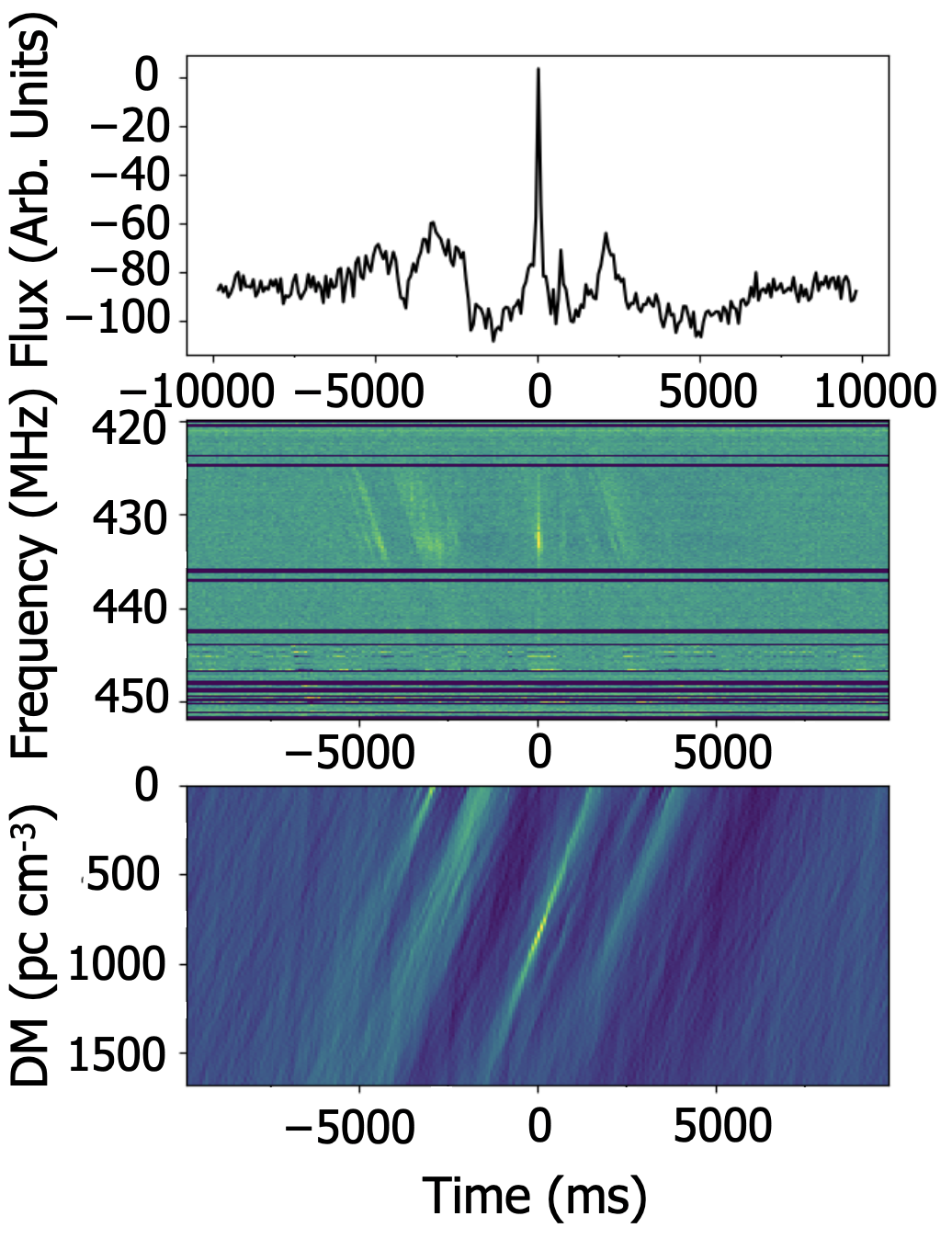}
\end{subfigure}
\caption{Detection plots of the four FRBs discovered in the survey. In each case, frequency channels corrupted by RFI have been masked (middle plots in each panel). Each panel shows the dedispersed pulse profile for the burst (top), signal strength (brightness) vs. frequency and time for the dedispersed pulse (middle), and signal strength (brightness) vs. DM and time (bottom).}
\label{fig-2}
\end{figure}

The remaining 4 events all had DMs significantly larger than the maximum Galactic DMs predicted by both the NE2001 \citep{cl02} and YMW16 \citep{ymw17} Galactic electron models. This suggests that they are FRBs, and Table \ref{tbl-1} shows these four FRBs and their parameters. Fig. \ref{fig-2} shows the detection plots, which show broadband signals and localization at non-zero DMs. The detection plots show a  vertical signal in the frequency band, indicating no channel delays after dedispersion has been applied. This dedispersion assumes a dispersion index of 2, corresponding to the expected (uncorrected) quadratic time delay as a function of frequency for cold plasma \citep{lk12}. There are no obvious visual deviations from this vertical morphology in the figure, indicating consistency with a dispersion index of 2. However, the signals are not sufficiently broadband to reliably fit for the dispersion index separately.

\subsection{Placement in the FRB Population}

Fig. \ref{fig-1} shows our four FRBs plotted with the currently known population of non-repeating FRBs taken from the FRBSTATS catalog \citep{frbstats}\footnote{\url{https://www.herta-experiment.org/frbstats/catalogue}}. Event data in this catalog were aggregated from several sources, including the Transient Name Server (TNS)\footnote{\url{https://www.wis-tns.org/}}, FRBCAT \citep{pbj+16}\footnote{\url{https://www.frbcat.org/}}, and the CHIME/FRB Catalogue \citep{chime21}\footnote{\url{https://www.chime-frb.ca/catalog}}.

As seen in Fig. \ref{fig-1}, there is a population of 10 FRBs detected by the Pushchino telescope (denoted by black dots) that have pulse widths that all exceed 300 ms \citep{fr19}. However, the observing bandwidth used in these detections was only 2.5 MHz, and only 6 frequency channels were used. This makes distinctions between RFI and dispersed astrophysical sources difficult (even at the low center frequency of 111 MHz that was used). Thus, it is uncertain if these are real sources or not, and we treat them as separate from the population of FRBs detected with other instruments (denoted by blue dots) in our analysis below. Apart these Pushchino detections, there are currently only four FRBs with measured pulse widths greater than 100 ms (all four were discovered by CHIME; \citealt{chime21}). 

We have used the Macquart relation \citep{mpm+20} to estimate a redshift for each of the four FRBs detected. The observed DM for an FRB can be broken up into four components: a contribution from the Galactic ISM, from the Galactic halo, from the intergalactic medium, and from the FRB host galaxy and any excess plasma local to the FRB. The DM from the Galactic ISM was estimated from the two Galactic electron models \citep{cl02,ymw17} (see Table \ref{tbl-1}). For the Galactic halo and host galaxy DM contributions, a value of 50 pc cm$^{-3}$ and $50/(1+z)$ pc cm$^{-3}$ have been assumed, respectively, in accordance with the literature \citep{pz19,mpm+20,jpm+21}. The resulting DM from the intergalactic medium was then converted to a redshift using the Macquart relation in Fig. 2 of \citet{mpm+20}. This figure indicates that there are uncertainties introduced from modeling and simulations (the shaded region of this figure). We have incorporated the uncertainties from the choice of Galactic electron model used and from the Macquart relation to estimate a redshift range for each of the FRBs. There are additional uncertainties in the assumed halo and host galaxy DM contributions which have not been incorporated in these estimates. The redshift ranges are presented in Table \ref{tbl-1}. 

One of our FRBs, FRB 920913, has a DM of 3338 pc cm$^{-3}$, the largest DM yet measured for an FRB (the next largest is FRB 20180906B, discovered with CHIME with a DM of 3038 pc cm$^{-3}$; \citealt{chime21}). \cite{jpm+21} indicate that the Macquart relation only applies up to some maximum redshift beyond which the trend reverses, so that larger DMs do not correspond to higher redshifts. For our unlocalized FRBs, this situation may apply, particularly for the high-DM FRB 920913. Thus our redshifts may be overestimates.

\subsection{Broadening from Interstellar Scattering}

The expected interstellar scattering times at 436 MHz for the four FRBs were determined from both the NE2001 and YMW16 models. Each 1 GHz scattering estimate from the models (assuming the maximum Galactic contribution along the line of sight) was scaled to the survey's center frequency of 436 MHz according to $\tau_{s} \sim f^{-4}$ (e.g., \citealt{okp+21} and references therein). Table \ref{tbl-1} shows these estimated scattering times.

All of the estimated scattering times are negligible (less than 2 ms) relative to the measured pulse widths, with the exception of FRB 910730. For this FRB, the Galactic scattering at 436 MHz is estimated to be 26 ms in YMW16 model, but only 0.03 ms in the NE2001 model. However, the YMW16 model does not use scattering as a modeling parameter. Instead, it estimates the scattering for a given DM value based on an  empirical scaling between scattering timescale and DM \citep{kmn+15}. We also see no indication of any one-sided asymmetric scattering tail in the pulse profile (Fig. \ref{fig-3}). Thus, the negligible scattering predicted by the NE2001 model is a more reliable indicator and is consistent with what we observe for FRB 910730. We therefore ignore scattering effects in our discussion since they are negligible.

\subsection{Broadening from Intra-channel Dispersion Smearing}

The DM smearing within frequency channels in the survey observations is 1.24 ms for every 100 pc cm$^{-3}$ of DM. We can estimate the effect that this has on pulse broadening by assuming that the observed (broadened) pulse is an intrinsic (Gaussian) pulse convolved with a Gaussian DM smearing function, so that the two contributions add in quadrature to produce the observed width. For the case of FRB 920913 (which has by far the largest DM of of our sample, 3338 pc cm$^{-3}$), the intra-channel smearing contributes 41 ms of broadening. For the observed pulse with of 157 ms, the deconvolved (intrinsic) pulse width would be about 151 ms, quite close to the observed width (less than a 4\% difference). For the other three FRBs, the pulse broadening is negligible (0.2\% or less of the deconvolved intrinsic pulse width). Thus, in all cases, the observed widths are good approximations (within a few percent) of the estimated intrinsic widths.

\subsection{Large Pulse Widths}

\begin{figure}
\includegraphics[width=\columnwidth]{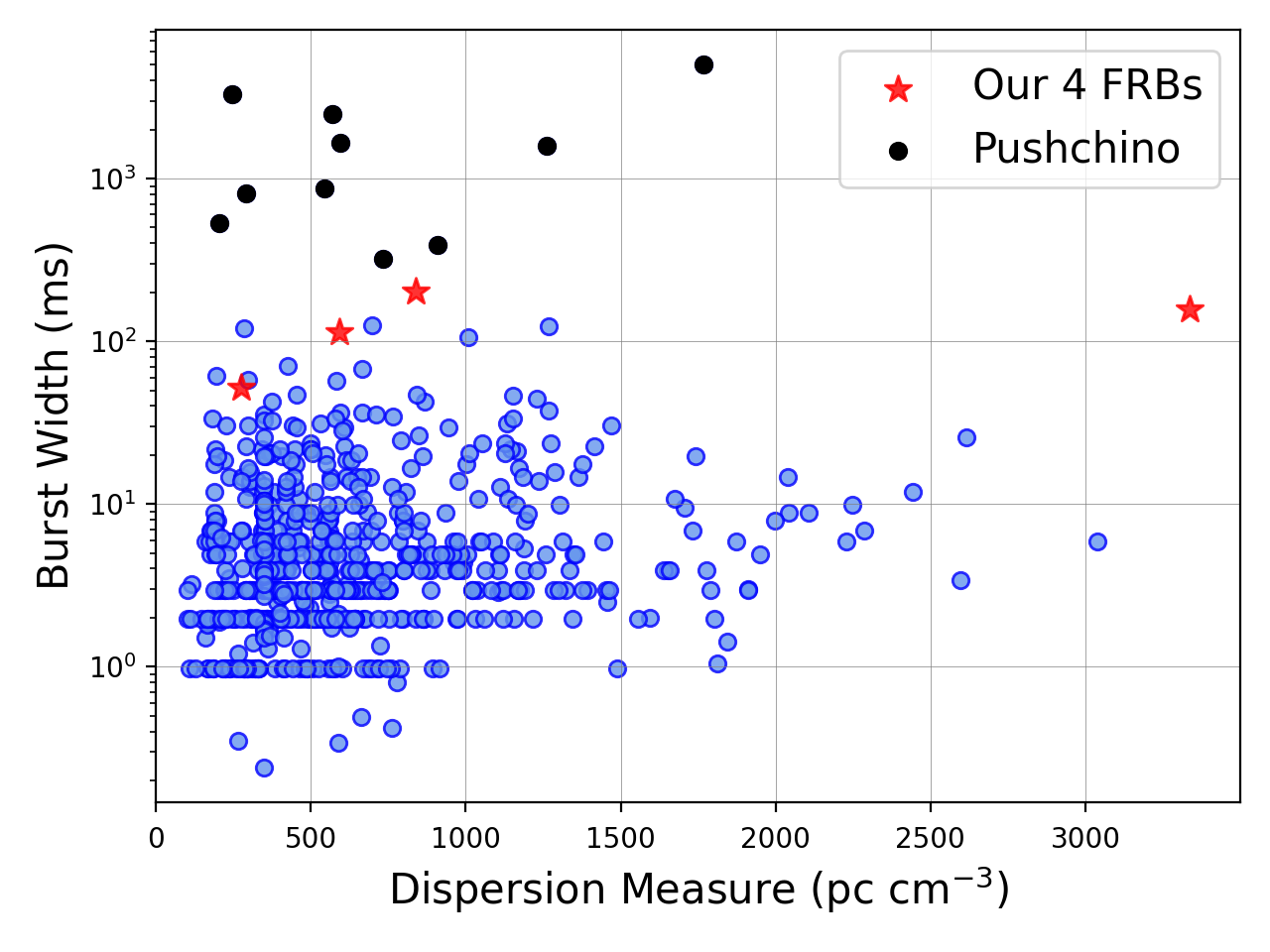}
\caption{Pulse width vs. DM for the catalog of currently known non-repeating FRBs (circles), plus the four new FRBs reported here (red stars). The subset of FRBs reported by \citet{fr19} from the Pushchino radio telescope have very large pulse widths and are shown as black circles. It is not clear of these are real detections of dispersed astrophysical signals or not (see comments in the text). Data for the plot were obtained from the FRBSTATS catalog \citep{frbstats}.}
\label{fig-1}
\end{figure}

\begin{figure}
\includegraphics[width=\columnwidth]{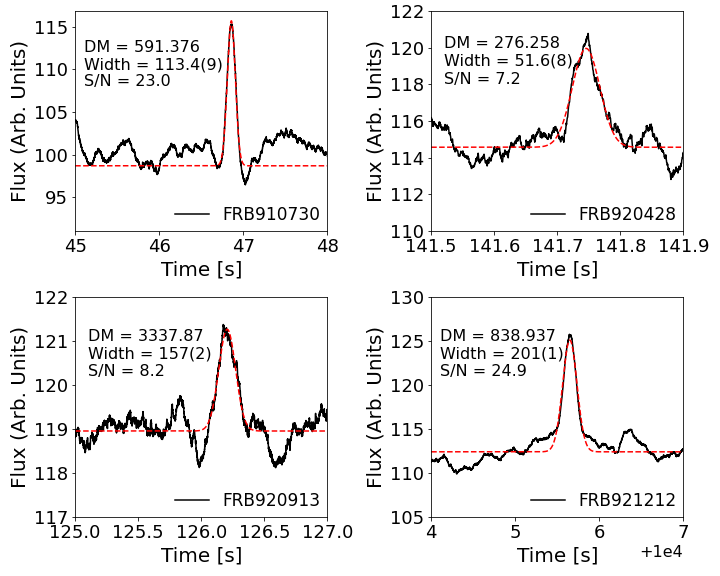}
    \caption{Dedispersed pulse profiles for the four FRBs reported here, using the RFI channel masking shown in Fig. \ref{fig-2}. A best-fit Gaussian in each case is indicated by the dashed red curve, and the resulting fit parameters are listed in Table \ref{tbl-1}. No clear evidence of a one-sided scattering tail is evident in any of the profiles. The horizontal axis represents the time after the start of the observation and the vertical axis is flux in arbitrary units. Neither axis is normalized to a single standard across the different panels, so comparisons between the different panels will not be to scale.}
    \label{fig-3}
\end{figure}

It is notable that three of the four FRBs that we have discovered have pulse widths above 100 ms, and the fourth FRB has a width of 52 ms (Table \ref{tbl-1}). As seen in Fig. \ref{fig-1}, these FRBs occupy a space in which such signals are rare.\footnote{See also Fig. 3 from \citet{phl+22}, where FRB detections are shown for different telescopes. Repeat bursts from FRB 20180916B (R3) that were detected with LOFAR at low frequencies are also shown here \citep{pmb+21}. Several of these repeater bursts exceed 100 ms in pulse width, but as noted by both \citet{pmb+21} and \citet{phl+22}, these events are likely dominated by scattering (unlike the FRBs reported here).} 

The vast majority of the FRB population has narrow pulse widths relative to our four detections. One question is why no such narrow-width FRBs were detected in our search given how much more commonly detected they are than wide-pulse FRBs. As noted above, we did find several relatively faint Galactic (i.e., smaller DM) signals with narrow widths ($< 4$ ms) that are probably RRATs, so our search was sensitive to such short-duration signals. However, the DMs of the four FRBs are an order of magnitude larger than these possible RRAT detections, and so the intra-channel DM smearing is also larger by this factor. For a typical FRB DM of 500 pc cm$^{-3}$, this contribution to pulse broadening would be about 7 ms, so FRBs with these DMs having widths less than this would be partially (or completely) smeared out. This would preclude detection of roughly half of the FRB population that has been cataloged to date (Fig. \ref{fig-1}) and could account for the lack of any detected FRBs with very narrow widths in our search.

Another factor leading to the preferential detection of FRBs with large pulse widths in our search may be the boxcar filter size used in HEIMDALL searches. We used a maximum window width of 512 samples, corresponding to 153 ms for our 0.3 ms sampled data. For more modern surveys with higher sampling rates ($< 0.1$ ms), this same maximum window size would be reduced in time accordingly, Thus, if other searches using HEIMDALL have not routinely used larger window sizes by default, they would not have had good sensitivity to FRBs with pulse widths $\ga 100$ ms. This would bias such searches against detection of wide pulses, leading to preferential detection of narrow-pulse FRBs, as seen in Fig. \ref{fig-1}.

\subsection{Flux Densities and Fluences}

\begin{figure}
\includegraphics[width=\columnwidth]{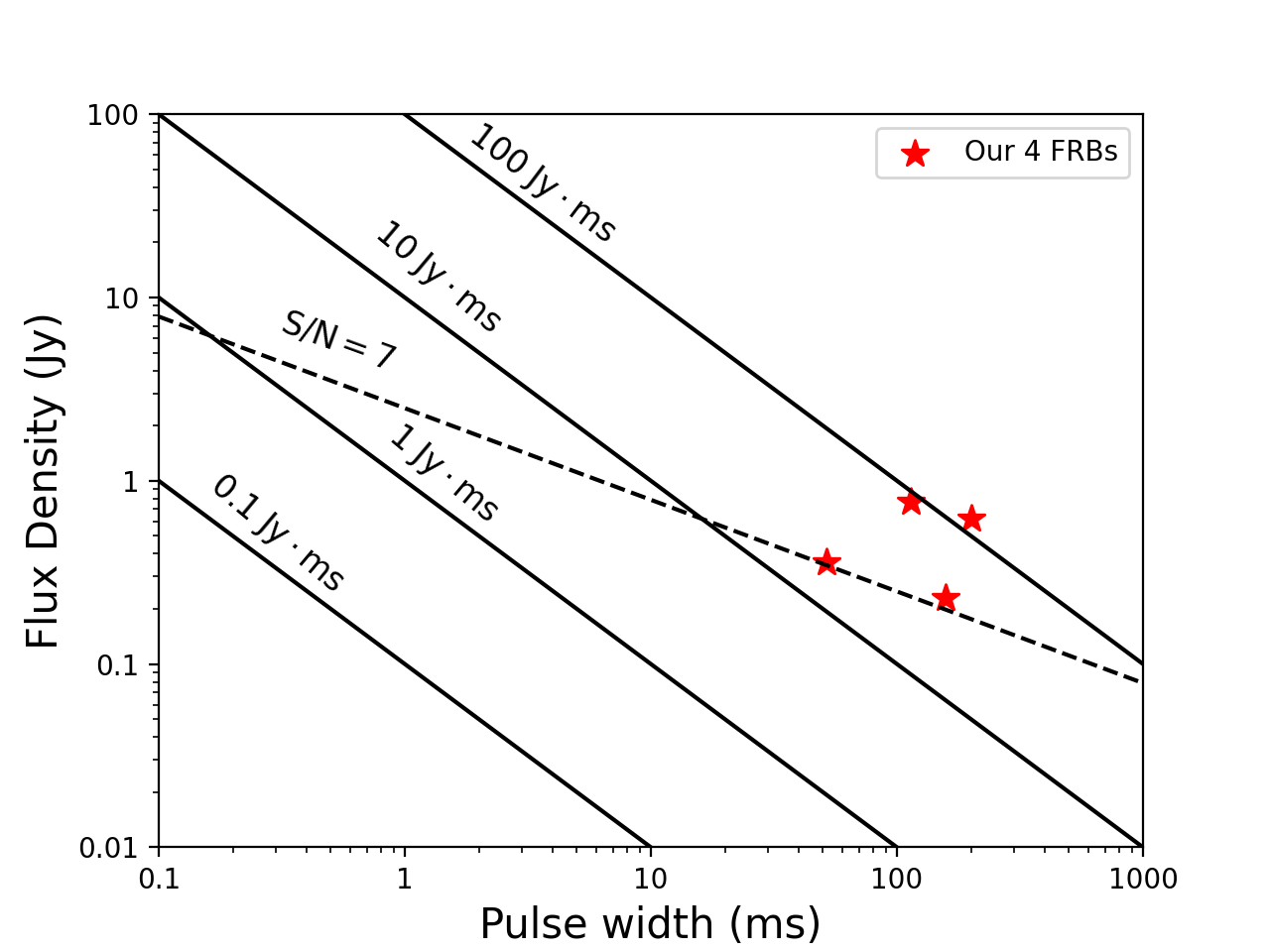}
    \caption{Flux density vs. pulse width parameter space for our survey analysis (see also \citealt{kp15}). Solid lines represent constant fluence values, and the dashed line represents our S/N detection threshold of 7. The four FRBs detected in the survey are indicated by red stars. For an assumed upper limit for FRB pulse widths of 200 ms, the fluence completeness is 35 Jy ms (see text).}
    \label{fig-4}
\end{figure}

The peak flux density $S$ in Table \ref{tbl-1} was calculated for each FRB detection by using the fitted parameters shown in Fig. \ref{fig-3} and the following expression (adapted from \citealt{rcl+13})

\begin{equation}
S = \frac{S_{sys} (S/N)}{W} \sqrt{\frac{W}{N_{p} \Delta f}}
\label{eqn-1}
\end{equation}

where $S_{sys}$ is the system noise of 90 Jy estimated for the survey \citep{mld+96}, $(S/N)$ is the measured FRB signal-to-noise ratio from the fitted amplitude and baseline noise level, $W$ is the measured pulse width, $N_{p} = 2$ is the number of polarizations recorded at the telescope, and $\Delta f = 32$ MHz is the observing bandwidth. The fluence $\mathcal{F}$ was then computed according to $\mathcal{F} = S W$. The fluence values for our FRBs range from 18 to 126 Jy ms, which is within the range of fluences observed for the FRB population to date. Fig. \ref{fig-4} shows the peak flux density vs. pulse width for our survey analysis (see also Fig. 2 of \citealt{kp15}). Lines of constant fluence (solid lines) and our detection S/N threshold of 7 (dashed line) are also shown, along with the four FRBs we have detected. For our completeness estimate, we assume that pulse widths will be less than 200 ms, slightly larger than the widest pulse we detected. The fluence threshold at 200 ms corresponding to our S/N detection limit of 7 is 0.17 Jy ms. This corresponds to a fluence completeness of 35 Jy ms. Although for smaller pulse widths we are sensitive to smaller fluences, we take this value to be our completeness threshold.

\subsection{Inferred All-sky FRB Rate}

We can use our FRB detections in this survey to calculate a corresponding daily all-sky event rate. We detected 4 events in the survey, where the survey spatial coverage can be approximated by the number of survey beams (43842) multiplied by the solid angle beam size for Parkes at 436 MHz ($1.35 \times 10^{-4}$ sr). This product is 5.92 sr, or 47\% of the full $4\pi$ sr sky. Each beam was nominally observed for 157 s, or $1.82 \times 10^{-3}$ days. The resulting nominal FRB detection rate is then $\mathcal{R}  = 4676$ events per day across the full sky. The fluence threshold for this rate can be approximated by our estimated fluence completeness of 35 Jy ms (for pulse widths less than 200 ms; see Fig. \ref{fig-4} and discussion above). 

Our inferred FRB all-sky rate above 35 Jy ms is roughly consistent (if no spectral index scaling is considered) with the values in Table 3 of \citet{phl+19}, which lists estimated 1400 MHz all-sky FRB rates that are typically in the range $\mathcal{R} \ge 10^{3}$ per day above a fluence threshold of a few Jy ms. However, converting those rates to the higher fluence threshold of 35 Jy ms
using the scaling expression $\mathcal{R} (> \mathcal{F}_{min}) \sim \mathcal{F}_{min}^{\gamma}$ with an assumed $\gamma = -1.5$ (from Euclidean geometry) \citep{phl+19} decreases these event rates significantly, resulting in just tens of events per day across the sky in most cases. For our survey, which covers about half of the sky but for only a small fraction of a day (see above), this would correspond to an expectation of order 0.1 FRBs detected in the survey (and in all cases less than 1).  

This number appears at first glance to be inconsistent with the four FRBs that we found, but the estimates listed in Table 3 of \citet{phl+19} were determined from FRB events that were in almost all cases much shorter in duration than ours (see the FRB population in Fig. \ref{fig-1}), and we found no such narrow-width FRBs. Thus our non-detection of any narrow-width FRBs is consistent with the expected number derived above. The fact that we detected 4 wide-pulse FRBs suggests that the rate of such wide FRBs may in fact be much larger than what might be expected from such rate estimates. This possibility remains to be explored and further studied.

\section{Conclusions}

We have discovered four new FRBs in the Parkes 70-cm pulsar survey data archive in a reprocessing of the data to look for dispersed single-pulses and bursts. The important results and conclusions from this work are as follows:

\begin{itemize} 

\item Each of the four FRBs discovered has a large pulse width ($> 100$ ms in three cases and 52 ms in the fourth case), which is not attributable to intra-channel dispersive smearing or Galactic scattering effects. These widths are significantly larger than the widths measured for almost all of the FRBs detected and cataloged to date (Fig. \ref{fig-1}). This may indicate that many more such signals could be present in pulsar surveys which could have been missed in searches that did not increase the search window to sufficiently large values. This possibility was hinted at by \citet{phl+22}, where the authors speculate that a population of ``not-so-fast radio bursts'' (nsFRBs) with durations of between 100 ms and several seconds could be waiting to be discovered.

\item One of the FRBs we discovered, FRB 920913, has a DM of 3338 pc cm$^{-3}$, which is the largest DM measured for any FRB detected and cataloged to date. 

\item All four of the FRBs were detected in survey observations that predate by almost a decade the observations in 2001 where first FRBs were detected and reported  (e.g., the Lorimer Burst; \citealt{lbm+07}). Thus, these four FRBs represent the first FRBs detected by any radio telescope so far (although they are of course not the first FRBs to be discovered and reported). 

\end{itemize}

The discoveries reported here illustrate the serendipitous nature of searching older, archival pulsar search data using newer techniques and wider parameter search ranges. This is made possible in part by increases in computing power and the availability of new search algorithms. Continued searches of archival pulsar surveys are likely to reveal more undiscovered FRBs in the future.

\section*{Acknowledgements}

We thank the anonymous referee for useful suggestions that have improved the manuscript. We also thank Emily Petroff and Michael Lam for useful discussions about the FRB population and pulse scattering. The Parkes radio telescope is part of the Australia Telescope National Facility (grid.421683.a), which is funded by the Australian Government for operation as a National Facility managed by CSIRO. We acknowledge the Wiradjuri people as the traditional owners of the Observatory site. F.C. is a member of the NANOGrav Physics Frontiers Center, which is supported by the NSF award PHY1430284. S.H. is supported by JSPS KAKENHI grant No. 20J20509 and the JSPS Overseas Challenge Program for Young Researchers.

\section*{Data Availability}

The raw data used in this study was provided by the CSIRO Data Access Portal (\url{https://data.csiro.au/}). Data products are available upon request to the corresponding author.

% FIGURES

% FIGURE: Waterfall plots

% FIGURE: Population plot
%%% width_dm_Pushchino.png has a label of Pushchino FRBs.%%%

% TABLES
% delete "*" in \begin{table*} and \end{table*} if you want to use this table in two columns
\begin{table*}
	\centering
	\caption{Measured properties of 4 FRBs discovered in the Parkes 70-cm pulsar survey archive. Values were determined from the Gaussian fit to each profile (Fig. \ref{fig-3}). The maximum Galactic DM contributions and scattering times listed are estimated from the NE2001 and YMW16 electron models, respectively. Scattering times $\tau{_s}$ have been scaled from 1 GHz to a center frequency of 436 MHz using the scaling relation $\tau_{s} \sim f^{-4}$. 
	The estimated redshift range $z$ was obtained from the Macquart relation \citep{mpm+20} and includes uncertainties from both the Galactic electron model used as well as the uncertainty in the relation seen in  Fig. 2 of \citet{mpm+20} (shaded region). In the case of FRB 920913, the large redshift from the large DM may be overestimated (see text discussion and \citealt{jpm+21}).}
	\setlength{\tabcolsep}{10pt}
	\begin{tabular}{lrrrr} 
		\hline
		%                           S02101_83           S17104_4        S22303_56       S32401_1
		FRB                         & 910730            & 920428        & 920913        & 921212 \\
		\hline
        Event MJD                   & 48467.934340      & 48740.759583  & 48878.035903  & 48968.257280    \\
		RA (J2000)                  & 07:06:45.9        & 17:09:00.0    & 15:03:00.0    & 21:46:15.0      \\
		DEC (J2000)                 & $-$43:33:00.0     & $-$15:36:00.0 & $-$05:12:00.0 & $-$07:47:00.0   \\
		FETCH Probability           & 0.9999354         & 0.99990344    & 0.99995935    & 0.99653363      \\
		S/N                         & 23.0              & 7.2           & 8.2           & 24.9            \\
		Width (ms)                  & 113.4(9)          & 51.6(8)       & 157(2)        & 201(1)          \\
		DM (pc cm$^{-3}$)           & 591.4             & 276.3         & 3337.9        & 838.9           \\
		DM$_{\rm Gal}$ (pc cm$^{-3}$)   & 136/251       & 160/126       & 35/31         & 42/30           \\
        $z$, redshift               & 0.16-0.53         & 0.02-0.05     & 2.03-4.64     & 0.45-1.04       \\
		$\tau_{s}$ (ms)             & 0.03/26.4         & 0.13/1.50     & 0.005/0.006   & 0.05/0.06       \\
		Flux Density (Jy)           & 0.77              & 0.36          & 0.23          & 0.62            \\
        Fluence (Jy ms)             & 87                & 18            & 37            & 126             \\
		\hline
	\end{tabular}
	\label{tbl-1}
\end{table*}

% REFERENCES
\bibliographystyle{mnras}
\bibliography{main} % if your bibtex file is called example.bib

% APPENDIX 
%\appendix

%\section{Some Extra Material}

%Extra material here if needed.

%%%%%%%%%%%%%%%%%%%%%%%%%%%%%%%%%%%%%%%%%%%%%%%%%%
% Don't change these lines
\bsp	% typesetting comment
\label{lastpage}
\end{document}